\documentclass[12pt]{article}
\usepackage[T1]{fontenc}
\usepackage{cite}
\usepackage{mathrsfs}
\usepackage{amsmath}
\usepackage{amssymb}
\usepackage[dvipdfm]{graphicx}
\usepackage{tabularx}
\usepackage{bm}

\newenvironment{minilinespace}{\baselineskip = 8mm}{}
\begin{document}

\begin{titlepage}

\begin{flushright}
{
	KUNS-2102
}
\end{flushright}
\vspace{1cm}

\begin{minilinespace}
\begin{center}
	{\Large
		{\bf A Note on Separability of Field Equations \\
            in Myers-Perry Spacetimes
 		}
	}
\end{center}
\end{minilinespace}
\vspace{1cm}

\begin{center}
Keiju Murata$^{1}$, Jiro Soda$^{2}$\\
\vspace{.5cm}
{\small {\textit{
Department of Physics, Kyoto University, Kyoto 606-8501, Japan
}}
}
\\
\vspace*{1.0cm}

{\small
{\tt{
\noindent
$^{1}$ murata@tap.scphys.kyoto-u.ac.jp
\\
$^{2}$ jiro@tap.scphys.kyoto-u.ac.jp
}}
}
\end{center}

\vspace*{1.0cm}



\begin{abstract}
We study separability of scalar, vector and tensor fields in 
5-dimensional Myers-Perry spacetimes with equal angular momenta. 
In these spacetimes, there exists enlarged symmetry, $U(2)\simeq SU(2)\times U(1)$. 
Using the group theoretical method with a twist,
we perform the dimensional reduction at the action level and
 show that both vector and tensor field equations can be reduced
  to coupled ordinary differential equations. 
 We reveal the structure of couplings between variables. 
In particular, we have obtained the decoupled master equations for 
 zero modes of a vector field. The same analysis can be done for zero modes of 
 a tensor field. Therefore, 
our formalism gives a basis for studying stability of Myers-Perry
black holes. 
\end{abstract}

\end{titlepage}

\section{Introduction}

Historically, black holes have played an important role to 
understand gravity. 
In the case of 4-dimensional Einstein gravity, there exist
Kerr black holes which are most general black holes in the vacuum.
There are many interesting features of Kerr black holes such as 
the existence of the horizon, Hawking radiation, and so on.  
However, for these properties to be relevant to reality,  
we have to prove the stability of Kerr black holes. 
Fortunately, it is proved that Kerr black holes are perturbatively stable.
It should be noted that the technique to show the stability is also useful to 
study the physics related to black holes, and hence gravity itself. 

Recently,  various higher dimensional black holes have been found
motivated by the development of string theory.
A straightforward generalization of Kerr black holes to higher dimensions
was obtained by Myers and Perry~\cite{Myers:1986un} many years ago.
These solutions have been further generalized to the cases with cosmological constant
and NUT parameters~\cite{Hawking:1998kw,
Gibbons:2004js,Gibbons:2004uw,Chen:2006xh,Hamamoto:2006zf},
the so-called Kerr-NUT-(A)dS spacetimes.  
Interestingly, the horizon with non-trivial topology is allowed
in higher dimensions. For example, black branes, black rings, etc are
 found~\cite{Horowitz:1991cd,Emparan:2001wn,Pomeransky:2006bd,Elvang:2007rd,Iguchi:2007is}. 
Needless to say,
it is important to study the stability of these black holes
in order to reveal the nature of higher dimensional gravity. 

To analyze the stability, it is important for field equations 
in black hole spacetimes to be separable and decoupled. 
 The separability is usually associated with spacetime symmetry. 
For example, consider the $d$-dimensional Schwarzschild spacetime
where $SO(d-1)$ is  symmetry of the spacetime. 
In this case, fields can be expanded in terms of $(d-2)$-dimensional spherical
harmonics. Since these harmonics constitute an irreducible representation of $SO(d-1)$, 
 the equations for each mode specified by eigenvalues
  are separated each other. In addition, the Schwarzschild spacetime is static, 
  hence these equations become ordinary differential equations. 
 In general, since a tensor field has components, we have to solve
 coupled ordinary differential equations. However, with a wise choice of variables,
 field equations in Schwarzschild spacetime turn out to be decoupled,
 namely, we have a single ordinary differential equation for each variable. 
Thus, in the case of higher dimensional Schwarzschild black holes,
field equations with any spin are
separable and decoupled~\cite{Kodama:2003jz,Ishibashi:2003ap,Kodama:2003kk}. 
The sufficient conditions for decoupling are not transparent contrary
 to the separability.  
 
In the case of Kerr black holes, the situation is more complicated.
Because of less symmetry, the separability of field equations is obscure.
Nevertheless, the Klein-Gordon equation in Kerr spacetime turns out to be separable. 
For vector and tensor fields, no one had expected the same miracle.
However, Teukolsky showed, in the null tetrad formalism, all of tensor field equations
can be separable and decoupled~\cite{Teukolsky:1973ha}. 
This result is highly non-trivial. In fact, the symmetry behind the separability
of fields in Kerr spacetime is not apparent. 

For higher dimensional rotating black hole spacetimes, 
there are many works concerning the separability of
geodesic Hamilton-Jacobi equations and Klein-Gordon equations. 
It turns out that these equations in Kerr-NUT-(A)dS spacetimes are 
separable because of the existence of a rank-2 closed conformal
 Killing-Yano tensor~
\cite{Frolov:2002xf,Frolov:2006ib,Kubiznak:2006kt,Page:2006ka,Frolov:2006pe,Krtous:2006qy,Krtous:2007xf,Houri:2007uq}. 
It is also shown that Kerr-NUT-(A)dS spacetimes are unique spacetimes which can
 accommodate a rank-2 closed conformal Killing-Yano tensor~\cite{Houri:2007xz}. 

The separability for vector and tensor field equations in higher dimensional
rotating black holes has not been well understood 
in contrast to the geodesic Hamilton-Jacobi equations and Klein-Gordon equations. 
In the case of Myers-Perry black holes with equal angular momenta and 
 dimensions higher than 6, 
it is shown that special components of a tensor field is
separable and decoupled~\cite{Kunduri:2006qa}. 
However, there was no separability analysis for general components even in those cases. 
In particular, to our best knowledge, there is no work on the separability
of vector and tensor field equations in 5-dimensional Myers-Perry black holes.
Here, we should stress that the issue of the separability
is generally different from the issue of the decoupling. 
In the case of geodesic Hamilton-Jacobi and Klein-Gordon equations,
the issue of decoupling is trivial. When we consider vector and  tensor fields,
we have to discriminate between separability and decoupling.

 In this paper, as a first step to study the stability of higher dimensional
 black holes, we study the separability of field equations in 
 5-dimensional Myers-Perry black holes with equal angular momenta. 
First of all, we should recall that
there is $U(1)\times U(1)$ symmetry in general 5-dimensional Myers-Perry black holes.
 The point is that, for equal angular momenta, the symmetry enhances to 
$U(2)\simeq SU(2)\times U(1)$. Using this enhanced symmetry and Hu's group theoretical 
method invented in the cosmological context~\cite{Hu:1974hh,Tanimoto:2004bc,Konoplya:1999kf}
with an important twist,
we will show separability of scalar, vector and
tensor fields. We do not insist this is a miracle like the Kerr case.
Rather, this is a consequence of the enhanced symmetry of the geometry.
However, in the coordinate basis, this separability is not apparent.
In this sense, the similarity exists 
between our analysis and that of Teukolsky. It should be stressed that
it is important to give a concrete analysis even in this special case.
Our work gives a basis for this direction. 

The organization of this paper is as follows. 
In Sec.\ref{sec:MPBH}, degenerate Myers-Perry black holes, i.e.,
5-dimensional Myers-Perry black holes with equal
angular momenta are reviewed. In particular, the symmetry of the spacetime is
discussed. 
In Sec.\ref{sec:Wigner}, we first review the group theoretical method proposed
 by Hu~\cite{Hu:1974hh}.
It is shown that Wigner functions constitute 
irreducible representations of the spacetime symmetry.
Next, we add a twist to Hu's method which allows us to prove the separability
of field equations in degenerate Myers-Perry black holes. 
In Sec.\ref{sec:scalar}, the separability of a scalar field is
shown. Although the result itself is well known~\cite{Frolov:2002xf},
 we rederive it using the group theoretical method. 
In Sec.\ref{sec:vector}, the separability of
a vector field  is proved. 
There, it turns out that the group theoretical method with the twist is essential.
We also reveal the structure of couplings between variables. 
In the subsection, we demonstrate the decoupling of equations for zero modes of a vector. 
The master equations for physical variables are obtained explicitly.
In Sec.\ref{sec:tensor}, the separability of a tensor field
 is shown. The result is important for the analysis of the stability of
 degenerate Myers-Perry black holes. 
The final section is devoted to discussions.

\section{Degenerate Myers-Perry black hole spacetimes}\label{sec:MPBH}

In this section, we introduce degenerate Myers-Perry black holes
and clarify the symmetry of the spacetime. 

Let us start with the 5-dimensional Myers-Perry black hole
\begin{eqnarray}
  ds^2&=& -dt^2 + \frac{\Sigma}{\Delta} dr^2 + \frac{\Sigma}{r^2} d\theta_1^2 
 +  \left( r^2 +a^2 \right) \sin^2 \theta_1 d\phi_1^2
 +  \left( r^2 +b^2 \right) \cos^2 \theta_1 d\phi_2^2  \nonumber \\
&& \quad +  \frac{m r^2}{\Sigma}(dt+a\sin^2 \theta_1 d\phi_1 
                                    + b\cos^2 \theta_1 d\phi_2 )^2 \ ,
\label{eq:MPBHG}
\end{eqnarray}
where
\begin{equation}
 \Sigma = r^2 \left( r^2 + a^2 \cos^2 \theta_1 + b^2 \sin^2 \theta_1  \right) \ , 
 \quad
 \Delta = (r^2+a^2)(r^2 +b^2) - m r^2\ ,
\end{equation}
and coordinate ranges are $0\leq \theta_1 < \pi /2$, 
$0\leq \phi_1 <2\pi$, $0\leq \phi_2 <2\pi$. 
This solution describes a general rotating black hole in 5-dimensions.
Parameters $a$ and $b$ characterize the rotation of the black hole.

Now, we consider the 5-dimensional Myers-Perry black hole with equal
angular momenta $a=b$. We call the resulting solution the degenerate Myers-Perry
black hole. 
Let us define new coordinates $\theta = 2\theta_1 $
$\phi =\phi_2 -\phi_1$ and $\psi =\phi_1 +\phi_2 $. 
Then, the metric is given by 
\begin{equation}
  ds^2= -dt^2 + \frac{\Sigma}{\Delta} dr^2 +
   \frac{r^2+a^2}{4}\left\{(\sigma^1)^2+(\sigma^2)^2+(\sigma^3)^2\right\} +
   \frac{m}{r^2+a^2}(dt + \frac{a}{2}\sigma^3)^2\ ,
\label{eq:MPBH}
\end{equation}
where $\Sigma$ and $\Delta$ reduce to
\begin{equation}
 \Sigma = r^2(r^2+a^2)\ ,\quad
 \Delta = (r^2+a^2)^2 - mr^2 \ .
\end{equation}
Here, we have defined the invariant forms $\sigma^a\,(a=1,2,3)$ of $SU(2)$ 
as
\begin{equation}
 \begin{split}
  \sigma^1 &= -\sin\psi d\theta + \cos\psi\sin\theta d\phi\ ,\\
  \sigma^2 &= \cos\psi d\theta + \sin\psi\sin\theta d\phi\ ,\\
  \sigma^3 &= d\psi + \cos\theta d\phi  \ ,
 \end{split}
\end{equation}
where  $0\leq \theta < \pi $, $0\leq \phi <2\pi$, $0\leq \psi <4\pi$.
It is easy to check the relation 
$d\sigma^a = 1/2 \epsilon^{abc} \sigma^b \wedge \sigma^c$.

Apparently, the metric~(\ref{eq:MPBH}) has the $SU(2)$ symmetry
characterized by Killing vectors $\xi_\alpha \ , (\alpha=x,y,z)$:
\begin{equation}
\begin{split}
 \xi_x &= \cos\phi\partial_\theta +
 \frac{\sin\phi}{\sin\theta}\partial_\psi -
 \cos\theta\sin\phi\partial_\phi\ ,\\
 \xi_y &= -\sin\phi\partial_\theta +
 \frac{\cos\phi}{\sin\theta}\partial_\psi -
 \cot\theta\cos\phi\partial_\phi\ ,\\
 \xi_z &= \partial_\phi\ .
\end{split}
\end{equation}
The symmetry can be explicitly shown by using
 the relation $\mathcal{L}_{\xi_\alpha}\sigma^a=0$,
 where $\mathcal{L}_{\xi_\alpha}$ is a Lie derivative
 along the curve generated by the vector field $\xi_\alpha$.
The dual vectors of $\sigma^a$ are given by
\begin{equation}
\begin{split}
 \bm{e}_1 &= -\sin\psi \partial_\theta +
  \frac{\cos\psi}{\sin\theta}\partial_\phi - \cot\theta\cos\psi
 \partial_\psi\ ,\\
 \bm{e}_2 &= \cos\psi \partial_\theta +
  \frac{\sin\psi}{\sin\theta}\partial_\phi - \cot\theta\sin\psi
 \partial_\psi\ ,\\
 \bm{e}_3 &= \partial_\psi\ ,
\end{split}
\end{equation}
and, by definition, they satisfy $\sigma^a_i\bm{e}^i_b = \delta^a_b$.  

From the metric (\ref{eq:MPBH}), we can also read off 
the additional $U(1)$ symmetry, which 
keeps the part of the metric, $\sigma_1^2+\sigma_2^2$. Thus, the symmetry of
5-dimensional degenerate Myers-Perry black hole 
becomes $SU(2)\times U(1) \simeq U(2)$
\footnote{The spacetime~(\ref{eq:MPBH}) also has  time translation
symmetry generated by $\partial/\partial t$. Due to this symmetry, we can
separate the  time dependence of fields as $\propto e^{-i\omega
t}$. However, this is obvious and we will not pay much attention to this
symmetry hereafter.}.  
The Killing vectors for the symmetry are 
$ \bm{e}_3 , \xi_x , \xi_y , \xi_z $, 
where $\bm{e}_3$ is a generator of $U(1)$ and $\xi_\alpha\,(\alpha=x,y,z)$
are generators of $SU(2)$. 
We will show the separability of field equations focusing on the symmetry.

For later calculations, it is convenient to define the new invariant forms
\begin{equation}
 \sigma^{\pm} = \frac{1}{2}(\sigma^1 \mp i \sigma^2)\ . 
\end{equation}
Here, we should notice the fact
\begin{eqnarray}
\mathcal{L}_{i e_3} \sigma^{\pm}
= i \partial_\psi \sigma^{\pm} =  \pm  \sigma^{\pm} \ .
 \label{rule}
\end{eqnarray}
The dual vectors for $\sigma^\pm$ are
\begin{equation}
 \bm{e}_{\pm} = \bm{e}_1 \pm i \bm{e}_2\ .
\end{equation}
By making use of these forms, the metric~(\ref{eq:MPBH}) can be rewritten as
\begin{equation} 
ds^2= -dt^2 + \frac{\Sigma}{\Delta} dr^2 +
   \frac{r^2+a^2}{4}\{4\sigma^+ \sigma^- + (\sigma^3)^2\} +
   \frac{m}{r^2+a^2}(dt + \frac{a}{2}\sigma^3)^2\ .
\label{eq:MPBH2}
\end{equation}

\section{Group Theoretical Method with a Twist }\label{sec:Wigner}

When we consider fields in the degenerate Myers-Perry spacetime~(\ref{eq:MPBH}), 
it is important to use a contrived expansion of fields 
with focusing the symmetry group, $SU(2)\times U(1)$. 
By doing so, we can show that any field equation in the degenerate
Myers-Perry spacetime can be separable. 
In this section, we first construct the irreducible representation of $SU(2)\times U(1)$
which is indispensable for the group theoretical method. 
Next, we explain our group theoretical method with a twist
used in this paper to show the separability.
It is the twist that allows us to prove the separability of field equations
in the degenerate Myers-Perry spacetime. 

Let us define  two kind of angular momentum operators
\begin{equation}
  L_\alpha = i \xi_\alpha \ , \quad
  W_a  = i \bm{e}_a \ .
\end{equation}
where $\alpha,\beta,\cdots = x,y,z$ and $a,b,\cdots = 1,2,3$. 
They satisfy commutation relations
\begin{equation}
 [L_\alpha, L_\beta] = i \epsilon_{\alpha\beta\gamma} L_\gamma\ ,
\end{equation}
and 
\begin{equation}
 [W_a, W_b] = -i\epsilon_{abc} W_c\ .
\end{equation}
They also commute each other $[L_\alpha, W_a]=0$. 
The symmetry group, $SU(2)\times U(1)$ are generated by $L_\alpha$ and
$W_3$. Note that $L^2 \equiv L_\alpha^2 = W_a^2$.

Let us construct the representation of  $U(2) \simeq SU(2)\times U(1)$.
The eigenfunctions of $L^2$ degenerate, but
can be completely specified by eigenvalues of other operators $L_z$ and $W_3$. 
The eigenfunctions are called Wigner functions defined by 
\begin{align}
 &L^2 D^J_{KM} = J(J+1)D^J_{KM}\ ,\label{eq:WigDef1}\\
 &L_z D^J_{KM} = M D^J_{KM}\ ,\label{eq:WigDef2}\\
 &W_3 D^J_{KM} = K D^J_{KM}\ ,\label{eq:WigDef3}
\end{align}
where $J,K,M$ are integers satisfying $J\geq 0,\  |K|\leq J,\  |M|\leq J$. 
From Eqs.~(\ref{eq:WigDef1}), (\ref{eq:WigDef2}) and (\ref{eq:WigDef3}), 
we see that $D^J_{KM}$ constitute the irreducible
representation of $SU(2) \times U(1)$.
The Wigner functions are functions of $(\theta,\phi,\psi)$ and
 satisfy the orthonormal relation
\begin{equation}
 \int^\pi_0 d\theta \int^{2\pi}_0 d\varphi \int^{4\pi}_0 d\psi
 \sin\theta\, D^J_{KM}(x^i)D^{J'\,\ast}_{K'M'}{}(x^i) =
 \delta_{JJ'}\delta_{KK'}\delta_{MM'}\ .
\end{equation}
The following relations are useful for later calculations
\begin{equation}
 W_+ D^J_{KM} = i \epsilon_K D^J_{K-1,M}\ ,\quad W_- D^J_{KM} = -i\epsilon_{K+1}D^J_{K+1,M}\
  , \quad W_3 D^J_{KM} = KD^J_{KM}\ ,
\end{equation}
where we have defined $W_\pm = W_1\pm i W_2$ and $\epsilon_K=\sqrt{(J+K)(J-K+1)}$. 
In a different notation, we have the following relations
\begin{equation}
\begin{split}
 \partial_+ D^J_{KM} &= \epsilon_K D^J_{K-1,M}\ ,\\
 \partial_- D^J_{KM} &= -\epsilon_{K+1}D^J_{K+1,M}\ ,\\
 \partial_3 D^J_{KM} &= -iKD^J_{KM}\ ,
\end{split}
\label{eq:delD}
\end{equation}
where we have defined 
$\partial_\pm \equiv e_\pm^i\partial_i$ and $ \partial_3 \equiv e_3^i\partial_i$.


Now, we are in a position to explain our approach to the analysis of
separability of field equations in degenerate Myers-Perry black holes. 
More than thirty years ago, Hu advocated the group theoretical method to study 
the dynamics of fields in the Bianchi type IX universe~\cite{Hu:1974hh}.
In the context of the Bianchi type IX cosmology, 
the expansion in terms of the Wigner functions
has been used to give a set of ordinary equations where
modes with different eigenvalue $K$ couple in general.
The group theoretical method is also extended to other Bianchi type
models (for instance, see \cite{Tanimoto:2004bc} and reference therein ).
It is our strategy to apply the group theoretical method to degenerate
Myers-Perry black holes. To do so, a twist is necessary.
The point is that the degenerate Myers-Perry solution has
the additional isometry $W_3$. Hence, modes with different $K$
should not mix  each other. 
As we will see soon, the scalar field is automatically separable
because of this reason. For vector and tensor fields, we have to
use the expansion in terms of the invariant basis $\sigma^a$
in addition to the expansion in terms of the Wigner functions $D^J_{KM}$. 
For example, when a vector $B_i $ is given, we have to expand as 
$B_i = \sigma_i^a B_a$.  Then, we can expand $B_a$ in terms of
the Wigner function. Here, we need
to take into account the fact that the invariant forms $\sigma^{\pm}$ carry 
the charge $\pm 1$ for $W_3$ as is shown in Eq.~(\ref{rule}). 
For example, $\sigma_i^{+} D^J_{K-1,M}$
has the eigenvalue $K$ for $W_3$. Therefore, when we consider the
eigenspace characterized by $K$, the correct expansion should be
\begin{eqnarray}
   B_i ( t,r , \theta, \phi ,\psi) &=& \sigma_i^{+} ( \theta, \phi ,\psi) 
   \sum_{JKM} B_{+}^{JKM} (t,r) D^J_{K-1 , M} ( \theta, \phi ,\psi) \nonumber\\
 && \quad + \sigma_i^{-} ( \theta, \phi ,\psi) 
   \sum_{JKM} B_{-}^{JKM} (t,r) D^J_{K+1 , M} ( \theta, \phi ,\psi) \nonumber\\
 && \quad + \sigma_i^{3} ( \theta, \phi ,\psi) 
   \sum_{JKM} B_{3}^{JKM} (t,r) D^J_{K , M} ( \theta, \phi ,\psi) \ ,
   \label{correct}
\end{eqnarray}
where the dependence on the coordinates is explicitly written.
The twist we used is to shift eigenvalues $K$ component by component.
Once we properly perform the expansion as is explained above,
any field equation in the degenerate 5-dimensional Myers-Perry black holes
will be separable. This is our main claim. 

Thus, in the group theoretical expansion method, we need to consider
fields in the tetrad basis. Moreover, we need a contrived expansion (\ref{correct}).
In this sense, the method is very similar to
the Teukolsky formalism. Of course, the decoupling of equations
is not guaranteed in general.

\section{Separability of scalar fields}\label{sec:scalar}

The separability of a scalar field in this spacetime~(\ref{eq:MPBH})
has been shown in~\cite{Frolov:2002xf}. However, we will show the separability
of the scalar field again with focusing on the spacetime symmetry 
$SU(2)\times U(1)$. We will show the separability at the action
level. This simplify the calculation.
The result will be useful when we  quantize the field. 

The action for a massive scalar field is given by
\begin{equation}
 S = -\frac{1}{2}\int d^5 x\sqrt{-g}\left[(\partial \phi)^2 +
		\mu^2\phi^2\right]\ ,
\label{eq:scalar_action}
\end{equation}
where $g_{\mu\nu}$ is the metric of the Myers-Perry black hole
and $\mu$ is the mass of the scalar field.  
Now, we transform basis vectors
$(\partial_\theta,\partial_\phi,\partial_\psi)$ to 
$(\bm{e}_{\pm}, \bm{e}_3)$. Then, the action for the scalar field
becomes
\begin{equation}
S = -\frac{1}{2}\int d^5
  x\sqrt{-g}\left[g^{AB}\partial_A \phi \partial_B \phi 
+ 2g^{Aa} \partial_A \phi \partial_a \phi 
+ g^{ab} \partial_a \phi \partial_b \phi
		+\mu^2\phi^2\right]\ .
\label{eq:scalar_action2}
\end{equation}
where $A,B\cdots= t,r$, $g^{Aa}\equiv g^{Ai}\sigma^a_i$, 
$g^{ab}\equiv g^{ij}\sigma^a_i\sigma^b_j$ and
$\partial_a \equiv e_a^i \partial_i$. 
The inverse metric in this frame reads
\begin{eqnarray}
&&  g^{tt} = - \frac{(r^2 +a^2)^2 + m a^2}{\Delta} \ ,\quad
  g^{rr} = \frac{\Delta}{\Sigma} \ ,\quad
  g^{t3} = \frac{2ma}{\Delta} \ , \nonumber\\
&&  g^{33} = \frac{4( r^2 + a^2 -m)}{\Delta} \ ,\quad
  g^{+-} = \frac{2}{r^2 + a^2} \ . 
\end{eqnarray}
We expand the scalar field $\phi$ by Wigner functions as, 
\begin{equation}
 \phi(x^\mu) = \sum_{JKM}\phi^{JKM}(x^A)D^J_{KM}(x^i)
 = \sum_{JKM}\phi^{*JKM}(x^A) D^{*J}_{KM}(x^i)\ .
\label{eq:scalar_dec}
\end{equation}
Substituting the expansion (\ref{eq:scalar_dec}) into the 
action~(\ref{eq:scalar_action}), we can carry out
($\theta,\varphi,\psi$) integration and the action becomes
\begin{equation}
 \begin{split}
  S = &-\frac{1}{16}\int dtdr \frac{\Sigma}{r} \sum_K\bigg[
-\frac{(r^2+a^2)^2+ma^2}{\Delta}|\dot{\phi}^K|^2 +
  \frac{\Delta}{\Sigma}|\phi^K{}'|^2\\ 
&+\left\{\frac{4(J(J+1)\Delta-K^2ma^2)}{\Delta(r^2+a^2)}+\mu^2\right\}|\phi^K|^2
+ \frac{2iKma}{\Delta}(\dot{\phi}^K\phi^K{}^\ast - \dot{\phi}^K{}^\ast\phi^K)
\bigg]\ .
 \end{split}
\label{eq:scalar_action3}
\end{equation}
where a dot and a prime denote the derivative with respect to $t$
and $r$, respectively. 
From Eq.~(\ref{eq:delD}), we see the operations of $\partial_\pm, \partial_3$
shift the eigenvalue $K$ of Wigner functions, 
but leave $J$ and $M$ invariant. Therefore, we have omitted the index $J$ and $M$
in the above action. We will omit $J$ and $M$ hereafter. 
It is remarkable that each mode with eigenvalues $J$ and $M$ is separated in the
action~(\ref{eq:scalar_action3}) because of $SU(2)$ symmetry, and 
each mode specified by $K$ is also separated because of $U(1)$ symmetry.

\section{Separability of vector fields}\label{sec:vector}

We consider a vector field in the background
metric~(\ref{eq:MPBH}). In the coordinate basis, it is difficult to
see the separability of field equations in contrast to the scalar field.
Hence, the group theoretical method plays an essential role here.  

The action for the vector field is given by
\begin{equation}
\begin{split}
 S &= -\frac{1}{4}\int d^5x \sqrt{-g}\, F_{\mu\nu}F^{\mu\nu} \ . 
\end{split}
\label{eq:EMaction}
\end{equation}
where $F_{\mu\nu} = \partial_\mu A_\nu - \partial_\nu A_\mu$
is the field strength. 
Let us expand the vector field $A_\mu$ in terms of invariant forms
and Wigner functions. 
First, consider components $A_t (x^\mu)$ and $A_r (x^\mu)$, 
which are scalar under the general covariant transformation
of $\theta,\varphi,\psi$ part.
The expansion by the Wigner functions gives
\begin{equation}
  A_t (x^\mu) = \sum_{K} A_t^{K}(x^A) D_{K}(x^i) \ , \quad
   A_r (x^\mu) = \sum_{K} A_r^{K}(x^A) D_{K}(x^i)\ .
\label{eq:AA_decom}
\end{equation}
Other components $A_i(x^\mu)$ $(i,j\cdots=\theta,\varphi,\psi)$, 
 behave as a vector under the general coordinate transformation
of $\theta,\varphi,\psi$ part. 
We need to find a basis which satisfies 
\begin{equation}
\begin{split}
 &L^2 D_{i,K}^a = J(J+1)D_{i,K}^a\ ,\\
 &L_z D_{i,K}^a = M D_{i,K}^a\ ,\\
 &W_3 D_{i,K}^a = K D_{i,K}^a\ ,
\end{split}
\label{eq:vec_Wigner}
\end{equation}
where operations are defined by Lie derivatives, that is 
$W_a D_{i,K}^b \equiv \mathcal{L}_{W_a} D_{i,K}^b$ and
$L_\alpha D_{i,K}^a \equiv \mathcal{L}_{L_\alpha} D_{i,K}^a$.
From Eq.~(\ref{eq:vec_Wigner}), 
we see that $D_{i,K}^a$ constitute the irreducible representation of 
$SU(2)\times U(1)$. 
Taking into account the property (\ref{rule}), 
we find  $D_{i,K}^a$ can be defined by
\begin{equation}
\begin{split}
 &D_{i,K}^+ = \sigma^+_i D_{K-1} \quad (|K-1|\leq J)\ ,\\
 &D_{i,K}^- = \sigma^-_i D_{K+1} \quad (|K+1|\leq J)\ ,\\
 &D_{i,K}^3 = \sigma^3_i D_{K} \qquad (|K|\leq J)\ .
\label{basis-v}
\end{split}
\end{equation}
This is nothing but the general rule (\ref{correct}).
 Using the basis (\ref{basis-v}), the field can be expanded as
\begin{equation}
 A_i(x^\mu) = \sum_K A_a^K(x^A) D_{i,K}^a(x^i)\ .
 \label{vector-expansion}
\end{equation}

Substituting the expansions (\ref{eq:AA_decom}) and (\ref{vector-expansion})
into the action (\ref{eq:EMaction}),
we obtain the dimensionally reduced action.
However, it is more convenient to transform basis vectors
$(\partial_\theta,\partial_\phi,\partial_\psi)$ to 
$(\bm{e}_{\pm}, \bm{e}_3)$ and change the components as
\begin{equation}
 A_a \equiv A_i e_a^i\ .
\end{equation}
It is easy to write down the expansion of $A_a$ in terms of Wigner functions: 
\begin{equation}
 \begin{split}
  &A_+(x^\mu) = \sum_K A_+^K(t,r) D_{K-1}(x^i)\ ,\\
  &A_-(x^\mu) = \sum_K A_-^K(t,r) D_{K+1}(x^i)\ ,\\
  &A_3(x^\mu) = \sum_K A_3^K(t,r) D_{K}(x^i)\ .
 \end{split}
\label{eq:Aa_dec}
\end{equation}
Notice that $A_+^K$, $A_-^K$ and $A_3^K$ are defined for
$|K-1|\leq J$, $|K+1|\leq J$ and $|K|\leq J$, respectively. 
We also define 
\begin{equation}
\begin{split}
 &F_{Aa} \equiv F_{Ai}e_a^i = \partial_A A_a - \partial_a A_A\ ,\\
 &F_{ab} \equiv F_{ij}e_a^i e_b^j = \mathcal{D}_a A_b - \mathcal{D}_b A_a\ ,
 \end{split}
\label{eq:F_e_basis}
\end{equation}
where we have defined the covariant derivative 
\begin{equation}
  \mathcal{D}_a A_b = \partial_a A_b - \omega_{ab}^c A_c \ , \quad
  \omega_{ab}^c = (e_{[a}^i \partial_{|i|} e_{b]}^j)\sigma_j^c \ .
\end{equation}
The explicit values of $\omega^c_{ab}$ are given by
\begin{equation}
 \omega^3_{+-}= -\omega^3_{-+} = i\ ,\quad
 \omega^+_{+3} = -\omega^+_{3+} = -\frac{i}{2}\ ,\quad
  \omega^-_{-3} = -\omega^-_{3-} = \frac{i}{2}\ , 
\end{equation}
and other components vanish. 
In the tetrad basis, the action for the vector field becomes
\begin{multline}
 S = -\frac{1}{4}\int d^5x \sqrt{-g}\,[g^{AB}g^{CD}F_{AC}F_{BD} +
 4g^{AB}g^{Ca}F_{AC}F_{Ba}
+ 2g^{Aa}g^{Bb}F_{AB}F_{ab}\\ + 2g^{AB}g^{ab}F_{Aa}F_{Bb} - 2g^{Aa}g^{Bb}F_{Ab}F_{Ba}
+ 4g^{Aa}g^{bc}F_{Ab}F_{ac} + g^{ab}g^{cd}F_{ac}F_{bd}]\ ,
\label{eq:EMaction1}
\end{multline}
where $g^{Aa}\equiv g^{Ai}\sigma^a_i$ and 
$g^{ab}\equiv g^{ij}\sigma^a_i\sigma^b_j$. 
Substituting Eq.~(\ref{eq:Aa_dec}) and (\ref{eq:F_e_basis}) into Eq.~(\ref{eq:EMaction1}), 
we get the effective 2-dimensional action for the vector field in
Myers-Perry black hole spacetime: 
\begin{equation}
 \begin{split}
  S = &-\frac{1}{8}\int dtdr \sum_K
  \bigg[-\frac{(r^2+a^2)^2+ma^2}{2r}|\dot{A}^K_r - A^K_t{}'|^2\\
&- \frac{2ma}{r}\text{Re}\left\{(\dot{A}^K_r -
  A^K_t{}')(A^K_3{}'+iKA^K_r)^\ast\right\}\\
& +\frac{\Delta}{r(r^2+a^2)}(|A^K_+{}'-\epsilon_K A^K_r|^2 + |A^K_-{}' +
  \epsilon_{K+1} A^K_r|^2)\\
&+ \frac{2(r^2+a^2-m)}{r}|A^K_3{}'+iKA^K_r|^2\\
& -\frac{r\{(r^2+a^2)^2+ma^2\}}{\Delta} (|\dot{A}^K_+ - \epsilon_K A^K_t|^2 + |\dot{A}^K_- +
  \epsilon_{K+1} A^K_t|^2)\\
&-\frac{2r(r^2+a^2)^2}{\Delta}|\dot{A}^K_3+iKA^K_t|^2\\
&-\frac{4mar}{\Delta}\text{Re}\big\{(\dot{A}^K_+ - \epsilon_K
  A^K_t)(\epsilon_K A^K_3 + iKA^K_+)^\ast \\
&\qquad\qquad+ (\dot{A}^K_- +
  \epsilon_{K+1}A^K_t)(-\epsilon_{K+1}A^K_3 + iKA^K_-)^\ast\big\}\\
&+\frac{2r}{r^2+a^2}|\epsilon_{K+1}A^K_- + \epsilon_K A^K_+ -2iA^K_3|^2\\
&+ \frac{4r(r^2+a^2-m)}{\Delta}(|\epsilon_K A^K_3 + iKA^K_+|^2 +
  |-\epsilon_{K+1} A^K_3 + iKA^K_-|^2)
\bigg]\ .
 \end{split}
\label{eq:EMaction2}
\end{equation}
Because of $U(2)$ symmetry, the equations for each mode specified by $J, K,M$ are 
separated each other. 

From the action (\ref{eq:EMaction2}), 
we can read off the structure of the couplings between variables. 
For example, it is easy to see $A_{\pm}$ of the zero modes are decoupled. 
The structure of the coupling can be displayed as follows:
\begin{equation}
\begin{array}{|c|c|c|}
   \multicolumn{3}{c}{J=0}   \\   \hline
A_+ & A_- & A_3, A_t , A_r \\ \hline
K=1   &       &              \\ \hline
      &       &  K=0         \\ \hline
      &  K=-1 &              \\ \hline
\end{array}
\nonumber
\end{equation}
In the above chart, variables in each row can couple to each other.
Apparently, $A_{\pm}$  are decoupled. 
We will present the zero mode analysis in the next subsection. 
From the action, we can also read off the couplings between variables for $J=1$.
\begin{equation}
\begin{array}{|c|c|c|}
   \multicolumn{3}{c}{J=1}   \\   \hline
A_+ & A_- & A_3, A_t , A_r         \\ \hline
K=2   &       &              \\ \hline
K=1   &       &  K=1         \\ \hline
K=0   &  K=0  &  K=0         \\ \hline
      &  K=-1 &  K=-1        \\ \hline
      &  K=-2 &              \\ \hline
\end{array}
\nonumber
\end{equation}
As one can see, the highest modes are always decoupled. 
We can continue these exercises for higher modes.
The similar rule also applies to tensor fields.

\subsection{Zero modes of Vector Field}

In this subsection, we study the zero modes $J=0$ of the vector field.  
As we will see soon, we have decoupled master equations for the vector zero modes.
This fact indicates the decoupling of other modes, 
although we could not prove it.

In this case, we need to consider only $A_t^0,A_r^0,A_+^1,A_-^{-1},A_3^0$.
Putting $J=M=0,K=\pm1$ in the action (\ref{eq:EMaction2}),
we see $A_{\pm}^{\pm 1}$ are decoupled from other components. 
 Thus, ignoring other components, 
we get the action for $A_{\pm}^{\pm 1}$, 
\begin{multline}
 S = \frac{1}{8}\int dt dr \bigg[
  \frac{r\{(r^2+a^2)^2+ma^2\}}{\Delta}|\dot{A}_\pm^{\pm 1}|^2 -
  \frac{\Delta}{r(r^2+a^2)}|A_\pm^{\pm 1}{}'|^2\\
 \mp
  \frac{2imar}{\Delta}(\dot{A}_\pm^{\pm 1} A_\pm^{\pm 1}{}^\ast -
  \dot{A}_\pm^{\pm 1}{}^\ast A_\pm^{\pm 1}) -
  \frac{4r(r^2+a^2-m)}{\Delta}|A_\pm^{\pm 1}|^2\bigg]\ .
  \label{zero-action}
\end{multline}
Equations of motion derived from the above action (\ref{zero-action}) are 
\begin{multline}
\left(\frac{\Delta}{r(r^2+a^2)}A_\pm^{\pm 1}{}'\right)' 
+ \frac{r}{\Delta}\left[\omega^2\{(r^2+a^2)^2+ma^2\} \mp 4\omega ma -
 4(r^2+a^2-m)\right]A_\pm^{\pm 1} = 0\ ,
\label{eq:EOM_K=pm1}
\end{multline}
where the time dependence is separated by moving on to the Fourier space, 
$A_\pm(t,r)=e^{-i\omega t}A_\pm(r)$. 
Now, we introduce a tortoise like coordinate $y_\ast$ by 
\begin{equation}
 dy_\ast \equiv \frac{r(r^2+a^2)}{\Delta}dr \ .
\end{equation}
%
%
%
Then, the equations of motions~(\ref{eq:EOM_K=pm1}) can be reduced to 
the Schr\"{o}dinger type:
\begin{equation}
 -\frac{d^2}{dy_\ast^2} A_\pm + V_\pm(r)  A_\pm = 0\ ,
\label{eq:Schro_pm}
\end{equation}
where the effective potentials read
\begin{eqnarray}
  V_\pm(r) = \frac{1}{r^2 +a^2}\left[
 4(r^2+a^2-m)-\omega^2\{(r^2+a^2)^2+ma^2\} \pm 4\omega ma 
                                          \right] \ .
\end{eqnarray}
%

%
%

The physical degrees of freedom of the massless vector in 5-dimensions
are three. Hence, we should have another wave equation. 
To extract the single equation from those for $A_t^0, A_r^0$ and $A_3^0$, 
we put $J=M=K=0$ in the action (\ref{eq:EMaction2}).
Then the action for $A_t^0, A_r^0$ and $A_3^0$ becomes
\begin{multline}
S = \frac{1}{8}\int dtdr \bigg[
\frac{(r^2+a^2)^2+ma^2}{2r}|\dot{A}_r^0-A_t^0{}'|^2 -
 \frac{2ma}{r}\text{Re}\{(\dot{A}_r^0 - A_t^0{}')A_3^0{}'{}^\ast\}\\
+\frac{2r(r^2+a^2)^2}{\Delta}|\dot{A}_3^0|^2-\frac{2(r^2+a^2-m)}{r}|A_3^0{}'|^2
 - \frac{8r}{r^2+a^2}|A_3^0|^2
\bigg]\ .
\label{eq:action_K=0}
\end{multline}
Now, we have to recall the gauge symmetry
\begin{equation}
 \begin{split}
  &A_t^0(t,r) \rightarrow A_t^0(t,r) - \partial_t \alpha(t,r)\ ,\\
  &A_r^0(t,r) \rightarrow A_r^0(t,r) - \partial_r \alpha(t,r)\ ,
 \end{split}
\end{equation}
where $\alpha(t,r)$ is an arbitrary function. Because of this gauge freedom
and a constraint equation, there remains only one physical degree of 
freedom in the
action~(\ref{eq:action_K=0}) which satisfies a single master equation. 
Let us show this explicitly. 
Equations of motion derived from the action~(\ref{eq:action_K=0}) are
\begin{align}
&\left(\frac{(r^2+a^2)^2+ma^2}{2r}(-i\omega A_r^0 - A_t^0{}')\right)' -
 \left(\frac{ma}{r}A_3^0{}'\right)' = 0\ ,\label{eq:delA_t}\\
&\frac{(r^2+a^2)^2+ma^2}{2r}(-i\omega A_r^0 - A_t^0{}') -
 \frac{ma}{r}A_3^0{}' = 0\ ,\label{eq:delA_r}\\
&\left(\frac{ma}{r}(-i\omega A_r^0 - A_t^0{}')\right)'
 +
 \left(\frac{2(r^2+a^2-m)}{r}A_3^0{}'\right)'\notag\\
 &\qquad\qquad\qquad    +
 2r\left(\frac{\omega^2(r^2+a^2)^2}{\Delta} -
 \frac{4}{r^2+a^2}\right)A_3^0 = 0  \ ,
 \label{eq:delA_3}
\end{align}
where the time dependence is separated in the Fourier space. 
Eq.~(\ref{eq:delA_t}) is nothing but the $r$ derivative of Eq.~(\ref{eq:delA_r}). 
From Eqs.~(\ref{eq:delA_r}) and (\ref{eq:delA_3}),
we can eliminate $-i\omega A_r^0 - A_t^0{}'$ to get the master
equation for $A_3^0$: 
\begin{equation}
 \left(\frac{(r^2+a^2)\Delta}{r\{(r^2+a^2)^2+ma^2\}}A_3^0{}'\right)' 
+ r\left(\frac{\omega^2(r^2+a^2)^2}{\Delta} -
 \frac{4}{r^2+a^2}\right)A_3^0 = 0\ .
\end{equation}
%
%
%
Using another tortoise like coordinate
\begin{equation}
 dw_\ast \equiv \frac{r\{ (r^2+a^2)^2+ma^2\}}{(r^2 +a^2)\Delta}dr \ ,
\end{equation}
we again get the Schr\"{o}dinger type equation
\begin{equation}
 -\frac{d^2}{dw_\ast^2}  A_3^0 + V_3(r)  A_3^0 = 0\ ,
\label{eq:Schro_3}
\end{equation}
where the effective potential reads
\begin{eqnarray}
    V_3(r) = \frac{(r^2+a^2)\Delta}{(r^2+a^2)^2+ma^2}
    \left(  \frac{4}{r^2+a^2} - \frac{\omega^2(r^2+a^2)^2}{\Delta} \right) \ .
\end{eqnarray}
Thus, it turns out that equations for zero modes $J=0$ reduce 
to  the decoupled ordinary differential equations (\ref{eq:Schro_pm}) 
and (\ref{eq:Schro_3}).

\section{Separability of tensor fields}\label{sec:tensor}

In this section, we discuss the separability of a tensor field $h_{\mu\nu}$
in the spacetime~(\ref{eq:MPBH}). The result is relevant to the stability analysis
of degenerate Myers-Perry black holes.
We will use units $16\pi G_5 =1$ in this section.

The action for the tensor field is 
\begin{multline}
 S = \frac{1}{4}\int d^5x \sqrt{-g}\,[-\nabla_\mu
 h_{\nu\rho} \nabla^\mu h^{\nu\rho} + \nabla_\mu h \nabla^\mu h \\
+2\nabla_\mu h_{\nu\rho} \nabla^\nu h^{\rho\mu} - 2\nabla^\mu h_{\mu\nu}
 \nabla^\nu h] \ ,
\label{eq:grav_action}
\end{multline}
where $\nabla_\mu$ denotes the covariant derivative with respect to $g_{\mu\nu}$
and we have defined $h= g^{\mu\nu} h_{\mu\nu}$.
The tensor field $h_{\mu\nu}$ can be classified into $h_{AB}$, $h_{Ai}$ and $h_{ij}$, 
which behave as scalar, vector and tensor under the general coordinate transformation
of $\theta,\varphi,\psi$ part, respectively. 
We already know how to expand the vector. 
As to the tensor, we need the basis $D_{ij,K}^{ab}$ which satisfy
\begin{equation}
\begin{split}
 &L^2 D_{ij,K}^{ab} = J(J+1)D_{ij,K}^{ab}\ ,\\
 &L_z D_{ij,K}^{ab} = M D_{ij,K}^{ab}\ ,\\
 &W_3 D_{ij,K}^{ab} = K D_{ij,K}^{ab}\ ,
\end{split}
\label{eq:tensor_Wigner}
\end{equation}
where operations are defined by Lie derivatives, that is 
$W_a D_{ij,K}^b \equiv \mathcal{L}_{W_a} D_{ij,K}^b$ and
$L_\alpha D_{ij,K}^a \equiv \mathcal{L}_{L_\alpha} D_{ij,K}^a$.
From Eq.~(\ref{eq:tensor_Wigner}), 
we see that $D_{ij,K}^a$ constitute the irreducible representation of 
$SU(2)\times U(1)$.
The general rule (\ref{correct}) tells us 
that $D_{ij,K}^{ab}$ can be defined by
\begin{equation}
\begin{split}
&D_{ij,K}^{++} = \sigma^+_i \sigma^+_j D_{K-2} \quad(|K-2|\leq J)\ ,\\
&D_{ij,K}^{+-} = \sigma^+_i \sigma^-_j D_{K} \qquad(|K|\leq J)\ ,\\
&D_{ij,K}^{+3} = \sigma^+_i \sigma^3_j D_{K-1} \quad(|K-1|\leq J)\ ,\\
&D_{ij,K}^{--} = \sigma^-_i \sigma^-_j D_{K+2} \quad(|K+2|\leq J)\ ,\\
&D_{ij,K}^{-3} = \sigma^-_i \sigma^3_j D_{K+1} \quad(|K+1|\leq J)\ ,\\
&D_{ij,K}^{33} = \sigma^3_i \sigma^3_j D_{K} \qquad(|K|\leq J)\ . \\
\end{split}
\end{equation}
Here, again, we have taken into account the property (\ref{rule})
so that the basis satisfy the relations (\ref{eq:tensor_Wigner}).
Thus, we can expand the tensor field as
\begin{equation}
\begin{split}
 &h_{AB}(x^\mu) = \sum_K h_{AB}^K(t,r) D_K(x^i)\ ,\\
 &h_{Ai}(x^\mu) = \sum_K h_{Aa}^K(t,r) D^a_{i,K}(x^i)\ ,\\
 &h_{ij}(x^\mu) = \sum_K h_{ab}^K(t,r) D^{ab}_{ij,K}(x^i)\ .
\end{split}
\label{eq:tensor_decom}
\end{equation}
If we substitute Eq.~(\ref{eq:tensor_decom}) into the action (\ref{eq:grav_action}),
 we can get the action for each mode labelled by $J$, $M$, $K$. 
With the same reason as scalar and vector fields, each eigen mode are
separated from others. Thus, the field equation for the tensor field 
is separable. 
Unfortunately, the result is messy and not so illuminated.
Hence, we refrain from displaying it.
We just note that, as is expected from the experience of the vector field,
we found some of the master equations for zero modes of the tensor field. 
 We hope to report our stability analysis
of degenerate Myers-Perry black holes in a separated
 paper~\cite{Murata}.

\section{Discussion}\label{sec:conc}

We have shown the separability of scalar, vector and tensor fields in the
5-dimensional Myers-Perry spacetime with equal angular momenta. This
spacetime has the enlarged symmetry $U(2)\simeq SU(2)\times U(1)$.
 We have focused on this symmetry and utilized the group theoretical method
 with the twist. More precisely, the fields are expanded by the invariant forms
 and the irreducible  representation of this symmetry with a shift of
 eigenvalues $K$ component by component. 
 As the result, each mode specified by eigenvalues has been separated 
 each other in the action. 
 The structure of couplings in the action is clarified. 
Most importantly, we have shown the existence of the master equation
for zero modes of vector and tensor fields. 
 In principle, we can extend our analysis to
higher dimensional rotating black holes by generalizing Wigner functions.
Of course, Kerr-AdS-NUT black holes can be investigated with the same strategy.
It is also intriguing to relate our approach to the Killing-Yano tensor.

The most important application of our results is to analyze
the stability of rotating black holes. 
In the case of Kerr-AdS black holes, the existence of superradiant
instability was discussed qualitatively based on AdS/CFT
correspondence~\cite{Hawking:1999dp}. 
This instability was investigated by using the gravitational perturbation in 
$(2N+5)$-dimensional Kerr-AdS black holes $(N\geq 1)$ with equal $(N+2)$ angular
momenta~\cite{Kunduri:2006qa}. 
In the case of Myers-Perry black holes, such a superradiant instability
has not been found. Instead, it has been argued that 
Myers-Perry black holes with sufficiently large angular momenta are
unstable~\cite{Emparan:2003sy}. 
Since we have master equations for zero modes,
 we will be able to analyze the stability of
degenerate Myers-Perry black holes. 

There are other 5-dimensional black holes which have $U(2)$ symmetry.  
Our analysis can be applicable to these black holes immediately.
One of them is a squashed black hole~\cite{Ishihara:2005dp}. 
This black hole looks like the 5-dimensional black hole in the vicinity
of the horizon, however, the spacetime far from the black hole
is locally that of the black string. It is well known that there exists Gregory-Laflamme
instability~\cite{Gregory:1993vy} in the black string. On the other hand, 5-dimensional
Schwarzschild black holes are stable. Therefore, 
it is interesting to study the stability of squashed black holes. 

It is also interesting to apply our method to toric Sasaki-Einstein manifold
$Y^{p,q}$ parametrized by two positive integers $p,q \ (p>q) $. 
The metric is given by
\begin{eqnarray}
  ds^2 = \frac{1-y}{6} \left[ (\sigma^1)^2 + (\sigma^2)^2 \right]
  + \frac{ dy^2}{w(y) q(y)} 
  + \frac{q(y)}{9} ( \sigma^3 )^2 
 + w(y) \left[ d\alpha + f(y)  \sigma^3  \right]^2
         \ , 
\end{eqnarray}
where
\begin{eqnarray}
&&  w(y) = \frac{2(b-y^2 )}{1-y} \ , \quad
  q(y) = \frac{b-3y^2 +2y^3}{b-y^2} \ , \quad
  f(y) = \frac{b-2y+y^2}{6(b-y^2)} \ , \nonumber \\
&&    b= \frac{1}{2} - \frac{p^2 - 3q^2}{4p^3} \sqrt{4p^2 - 3q^2} \ .
\end{eqnarray}
The coordinates have the following range:
\begin{eqnarray}
y_1 \leq y \leq y_2 \ , 0\leq \theta \leq \pi \ , 0\leq \phi \leq 2\pi \ ,
0\leq \psi \leq 2\pi \ , 0 \leq \alpha \leq 2\pi \ell \ , 
\end{eqnarray}
where
\begin{eqnarray}
 y_{1,2} = \frac{1}{4p} \left( 2p\mp 3q -\sqrt{4p^2 -3q^2} \right) \ , 
 \ell = \frac{q}{3q^2 -2p^2 +p \sqrt{4p^2 -3 q^2 }} \ .
\end{eqnarray}
The point is that the symmetry of this manifold is nothing but $U(2)$.
Hence, the method we have explained in this paper can be applicable. 
The spectrum of a scalar Laplacian in this manifold have been already 
studied in~\cite{Kihara:2005nt}. However, the spectrum of vector and
 tensor Laplacians have not been
examined yet and can be investigated using our method. 
It would be important for the AdS/CFT correspondence.

\section*{Acknowledgements}
K.M. is supported in part by JSPS Grant-in-Aid for Scientific Research,
No.193715 and also by the 21COE program ``Center for Diversity and
Universality in Physics,'' Kyoto University. 
J.S. is supported by  
the Japan-U.K. Research Cooperative Program, the Japan-France Research
Cooperative Program,  Grant-in-Aid for  Scientific
Research Fund of the Ministry of Education, Science and Culture of Japan 
 No.18540262 and No.17340075. 



\bibliographystyle{kuma}
\bibliography{separability_and_MPBH}

\end{document}